\documentclass[usenatbib]{mn2e}
\usepackage{epsfig}
\def\Xeii{\hbox{Xe\,{\sc ii}}}
\def\Pii{\hbox{P\,{\sc ii}}}
\def\Srii{\hbox{Sr\,{\sc ii}}}
\def\Feii{\hbox{Fe\,{\sc ii}}}
\def\Gaii{\hbox{Ga\,{\sc ii}}}
\def\Krii{\hbox{Kr\,{\sc ii}}}
\def\Ptii{\hbox{Pt\,{\sc ii}}}
\def\Auii{\hbox{Au\,{\sc ii}}}
\def\Hgii{\hbox{Hg\,{\sc ii}}}
\def\Tlii{\hbox{Tl\,{\sc ii}}}
\def\Biii{\hbox{Bi\,{\sc ii}}}
\newcommand{\vsini}{\mbox{$v\sin\!i$}}
\newcommand{\logg}{\mbox{$\log g$}}
\newcommand{\teff}{\mbox{$T_{\mathrm{eff}}$}}
\newcommand{\vturb}{$\xi$}
\newcommand{\loggf}{\mbox{log {\em gf}}}

\begin{document}

\title[Xenon in HgMn Stars]{Xenon in Mercury-Manganese Stars}
\author[M.M. Dworetsky, J.L. Persaud \& K. Patel]
{M.M. Dworetsky$^{1}$\thanks{E-mail: mmd@star.ucl.ac.uk (MMD); 
jlp@star.ucl.ac.uk (JLP)}, J. L. Persaud$^{1}$\footnotemark[1] and K.
Patel$^{1}$\\
$^{1}$Dept. of Physics \& Astronomy, University College London,
Gower Street, London WCIE 6BT, United Kingdom}

\date{Accepted 2008 January 8.  Received 2007 December 
18; in original form 2007 August 20}

\maketitle
\begin{abstract}
Previous studies of elemental abundances in Mercury-Manganese (HgMn) 
stars have occasionally reported the presence of lines of the ionized 
rare noble gas \Xeii,\ especially in a few of the hottest stars with 
\mbox{$\teff\sim13000$--15000 K}. A new study of this element has been 
undertaken using observations from Lick Observatory's Hamilton 
\'{E}chelle Spectrograph. In this work, the spectrum synthesis program 
UCLSYN has been used to undertake abundance analysis assuming LTE. We 
find that in the Smith \& Dworetsky sample of HgMn stars, Xe is vastly 
over-abundant in 21 of 22 HgMn stars studied, by factors of 3.1--4.8 
dex. There does not appear to be a significant correlation of Xe 
abundance with \teff. A comparison sample of normal late B stars shows 
no sign of \Xeii\ lines that could be detected, consistent with the 
expected weakness of lines at normal abundance. The main reason for the 
previous lack of widespread detection in HgMn stars is probably due to 
the strongest lines being at longer wavelengths than the photographic 
blue. The lines used in this work were $\lambda$4603.03, 
$\lambda$4844.33 and $\lambda$5292.22.
\end{abstract}


\section{Introduction}

Among the chemically peculiar (CP) stars of the upper main sequence, the
HgMn stars \mbox{(10500 K $\leq \teff \leq 14500$ K)} constitute some of the
objects most intensively studied with a wide range of optical and
ultraviolet spectra (summarised by \citealt{wal04}). The reasons for this
include that: many of these stars are extremely slow rotators, making
details of their line profiles easily studied for isotopic anomalies, for
example; they are not affected by strong magnetic fields, which simplifies
theoretical considerations; and there are many of these stars bright enough
to study with the highest spectral resolutions available. Most often these
stars have been studied individually, but a few studies have considered
single elements among a set of stars to determine the overall behaviour of
abundances and isotope distributions as a function of physical variables
such as \teff\ (for example, \citealt{s93,s94,s96,s97,sd93,wl99,db2000}).
Here, we report an analysis of strong unblended lines of the rare noble gas
xenon (atomic number 54; cosmic \mbox{log {\em A}\,(Xe) = 2.27} on the scale
where \mbox{log {\em A}\,(H) = 12.00}) with a view to determining its
abundance in a sample of HgMn stars. As will be shown, Xe is found in nearly
all HgMn stars and exhibits abundance excesses of 3 dex or more.

Previous observations of \Xeii\ in the spectra of HgMn stars are scanty,
although it was originally reported as a possible identification in 3 Cen
A by \citet{bid62} and a probable identification was reported by
\citet*{andersen_etal84} in the He-weak Bp star HR\,6000 which seems to be
closely related to the HgMn stars. More recently, \Xeii\ was identified in
HR 7361 by \citet{adel92}; 46 Aql by \citet*{sada-etal01}; $\kappa$ Cnc by
\citet{rs88}; 33 Gem by \citet*{adel-etal96}; 112 Her by
\citet*{ryab-etal96}. This sparseness may be attributable to the past
tendency to observe these stars mainly in the blue photographic region,
where \Xeii\ lines are relatively weak. For example, \citet{guth85}
explored Cowley's Dominion Astrophysical Observatory 2.4\,\AA\,mm$^{-1}$
spectrograms of several HgMn stars but did not report \Xeii.  In this
work, the strong lines of \Xeii\ used were $\lambda$5292.22 and
$\lambda$4844.33, supplemented by the weaker $\lambda$4603.03 visible in
most of the programme stars.

\section{observations}

The list of stars analysed for this study is from the paper of
\citet{sd93} who originally listed 26 stars. In Table~\ref{sample}, two
stars are omitted that are not HgMn stars, 36 Lyn (a magnetic Ap star)
and HR 6000, a hot analogue of the HgMn stars. Two other stars were also
omitted, $\varphi$ Phe and $\nu$ Cnc, due to absence of good spectra.
This left a sample of 22 HgMn stars.

Apart from $\beta$\,Scl, the observations for this study were obtained
with the Hamilton \'{E}chelle Spectrograph (HES; \citealt{vogt87,mis97})
at Lick Observatory, fed by the 0.6-m Coud\'{e} Auxilliary Telescope,
during four observing runs in 1994--1997. Prior to our 1994 observations,
improvements were made in the resolution and instrumental profile of the
HES by the replacement of some of the optical components, which also made
it possible to use the full field of the $2048 \times 2048$ CCDs to
maximum advantage. For our observations, both the unthinned
phosphor-coated Orbit CCD (Dewar 13) and, from July 1995, the thinned Ford
CCD (Dewar 6) were used, depending on availability as the latter was
shared with the multi-object spectrograph on the 3-m telescope. The Orbit
CCD has very few bad pixels or columns.  The Ford CCD has several column
defects but it has a much higher quantum efficiency in the blue and so we
used the Ford CCD whenever it was available. The spectral range for the
observations was 3800--9000\AA\ and the typical signal-to-noise ratio
(S/N) per pixel in the centres of orders ranged from 75 to 250 depending
on wavelength. With the slit-settings used, the combination of
spectrograph and CCDs gave resolutions $R \approx 46\,500$. We used the
polar axis quartz lamp for flat fields and a Th--Ar comparison for
wavelength calibration.

The observations of $\beta$\,Scl are from the UVES Paranal Observatory 
Project (ESO DDT Program ID 266.D-5655) which is an on-going program to 
observe, reduce and provide a public library of good quality spectra of 
bright southern stars obtained with the Ultraviolet and Visual 
\'{E}chelle Spectrograph (UVES) mounted on the Very Large Telescope 
(VLT) unit Keuyen (VLT UT2). The UVES Paranal Observatory Project is 
described by \citet{bag-etal03}. The project provides almost complete 
wavelength coverage from 3000--10\,000\AA\ for all stars observed. The 
spectral resolution is about 80\,000 and the typical S/N is 300--500 in 
the \mbox{V-band}.

The Lick Observatory \'{E}chelle spectra were extracted and calibrated
using standard {\sc iraf} extraction packages \citep{val90,church95},
running on the Starlink node of University College London (UCL). 
Previous measurements \citep{al98} showed that there were no measurable
effects of parasitic light (residual scattered light) in the line
profiles provided that general scattered light in the adjacent
interorder spaces was taken as the subtracted background. In practice,
the residual scattered light was less than approximately 1 per cent and
so we have made no corrections for it (see \citealt{db2000} for further
details).
 
\begin{table*}
\begin{minipage}{12cm}

\caption{Programme HgMn stars}

\label{sample}
\begin{tabular}{lrlcccccc}

\hline

Star & HD & \teff & \logg & Ref. & \vturb &  Ref. & \vsini & Ref.\\
 & & (K) & (cgs) & & (km\,s$^{-1}$) & & (km\,s$^{-1}$) & \\

\hline

87\,Psc & 7374 & 13150 & 4.00 & (1) & 1.5 & (1) & 21 & (4)\\
53\,Tau & 27295 & 12000 & 4.25 & (1) & 0.0 & (2) & 5 & (3, 5)\\
$\mu$\,Lep & 33904 & 12800 & 3.85 & (1) & 0.0 & (2) & 15.5 & (4)\\
HR\,1800$^a$ & 35548 & 11050 & 3.80 & (1) & 0.5 & (1) & 3 & (4)\\
33\,Gem & 49606 & 14400 & 3.85 & (1) & 0.5: & (1) & 19.5 & (5)\\
HR\,2676 & 53929 & 14050 & 3.60 & (1) & 1.0: & (1) & 21 & (5)\\
HR\,2844 & 58661 & 13460 & 3.80 & (1) & 0.5: & (1) & 27 & (4)\\
$\kappa$\,Cnc$^a$ & 78316 & 13200 & 3.7 & (6) & 0.0 & (2) & 6 & (3)\\
HR\,4072$^a$ & 89822 & 10650 & 3.8 & (7) & 1.0 & (7) & 3.2 & (7)\\
$\chi$\,Lup$^a$ & 141556 & 10650 & 3.90 & (8) & 0.0 & (8) & 2 & (7)\\
$\iota$\,CrB$^a$ & 143807 & 11000 & 4.00 & (1) & 0.2 & (9) & 1 & (7)\\
$\upsilon$\,Her & 144206 & 12000 & 3.80 & (1) & 0.6 & (10) & 9 & (4)\\
$\phi$\,Her$^a$ & 145389 & 11525 & 4.05 & (11) & 0.4 & (11) & 8.0 & 
(11)\\
28\,Her & 149121 & 11000 & 3.80 & (1) & 0.0 & (12) & 8 & (12)\\
HR\,6997 & 172044 & 14500 & 3.90 & (1) & 1.5 & (1) & 34 & (1)\\
112\,Her$^a$ & 174933 & 13100 & 4.10 & (13) & 0.0 & (13) & 6 & (13)\\
HR\,7143 & 175640 & 12100 & 4.00 & (1) & 1.0 & (1) & 2 & (4)\\
HR\,7361 & 182308 & 13650 & 3.55 & (1) & 0.0 & (1) & 9 & (1)\\
46\,Aql & 186122 & 13000 & 3.65 & (1) & 0.0 & (1) & 1 & (1)\\
HR\,7664 & 190229 & 13200 & 3.60 & (1) & 0.8 & (12) & 8 & (4)\\
HR\,7775 & 193452 & 10800 & 3.95 & (1) & 0.0 & (1) & 1 & (1)\\
$\beta$\,Scl & 221507 & 12400 & 3.90 & (1) & 0.0: & (1) & 25 & (4)\\

\hline

\end{tabular}

\begin{minipage}{12cm}
$^a$Binaries with two spectra. Values given for \teff, \logg,\ \vsini\
and microturbulent velocity, \vturb, are for the primary star. 

Values of \vturb\ followed by a colon (:) are approximate and were 
derived solely from ultraviolet \Feii\ lines by (1).

References: (1) \citet{sd93}; (2) \citet{adel88a}; (3) \citet{adel87}; 
(4) \citet*{djs98}; (5) Determined from this work; (6) 
\citet{ryab_etal98}; (7) \citet{har97}; (8) \citet*{wal_etal94}; (9) 
\citet{adel89}; (10) \citet{adel_fuhr85}; (11) \citet{zav2007}; (12) 
\citet{adel88b}; (13) \citet{ryab-etal96}.

\end{minipage}
\end{minipage}

\end{table*}

\section{abundance determination}
\subsection{Stellar parameters and stellar atmospheres}

The effective temperatures and surface gravities of the HgMn programme
stars are given in Table~\ref{sample}. Seven of the stars are listed as
being binaries with two spectra; of these, six are noted as double-lined
spectroscopic binaries but HR\,1800 is more accurately described as a
close visual binary where the secondary spectrum is evident as
rotationally-broadened features. The adopted stellar data and light
ratios for the binary stars are shown in Table~\ref{binaries}. With the
exception of $\phi$\,Her, the light ratios follow those used in our
previous work \citep{db2000}. We obtained the light ratio for
$\phi$\,Her by using the magnitude difference of 2.57 magnitudes at
5500\AA\ as determined by \citet{zav2007}. Light ratios were
interpolated to the relevant wavelengths for the \Xeii\ lines in this
study using the \cite{kurucz93} model atmosphere fluxes. Although
HR\,7775 is known to have a faint visual secondary contributing $\sim$
2--4 per cent of the flux (\citealt*{boh_etal98}), in this analysis we
have treated it as a single star and ignored the presence of the
secondary.

In Table~\ref{sample}, when \vsini\ is listed as being determined from
this work it was calculated using the spectrum synthesis program UCLSYN
\citep{sd88,s92} which calculates a rotational profile from the
limb-darkening and then performs a $\chi^{2}$ minimization to get a best
fit, holding other parameters constant.  This was done for several lines
in the spectrum and then the average \vsini\ was adopted.

\begin{table*}
\begin{minipage}{13.2cm}

\caption{Binary stars: adopted stellar data and light ratios.}

\label{binaries}

\begin{tabular}{lccllcccc}

\hline

Star & $\lambda$ & $L_{\mathrm{A}}/L_{\mathrm{B}}$ &
$T_{\mathrm{effA}}/\log g_{\mathrm{A}}$ & $T_{\mathrm{effB}}/\log 
g_{\mathrm{B}}$ & $L_{\mathrm{A}}/L_{\mathrm{B}}$ & 
$L_{\mathrm{A}}/L_{\mathrm{B}}$ & $L_{\mathrm{A}}/L_{\mathrm{B}}$ 
& Ref.\\
 & & & (K)/(cgs) & (K)/(cgs) & 4603\,\AA & 4844\,\AA & 5292\,\AA\\

\hline

HR\,1800 & $H_{p}$ & 2.45 & 11050/3.8 & 9500/4.0 & 2.44 & 2.63 & 
2.39 & (1)\\
$\kappa$\,Cnc & 5480 & 11.5 & 13200/3.7 & 8500/4.0 & 12.7 & 14.3 & 
11.8 & (2)\\
HR\,4072 & 4520 & 5.45 & 10650/3.8 & 8800/4.2 & 5.39 & 5.77 & 5.17 & 
(3)\\
$\chi$\,Lup & 4520 & 3.65 & 10650/3.9 & 9200/4.2 & 3.64 & 3.87 & 3.56 
&(4)\\
$\iota$\,CrB & 4520 & 2.70 & 11000/4.0 & 9000/4.3 & 2.67 & 3.05 & 2.56 & 
(3)\\
$\phi$\,Her & 5500 & 10.67 & 11525/4.05 & 8000/4.30 & 12.45 & 12.70 & 
11.05 & (5)\\
112\,Her & 4520 & 6.20 & 13100/4.1 & 8500/4.2 & 6.14 & 6.94 & 5.70 & 
(6)\\

\hline

\end{tabular}

\begin{minipage}{13.2cm}
Note: The entry for HR\,1800 is the ratio quoted for the broadband 
$H_{p}$ filter, which we assume to be the light ratio at 
\mbox{H$\beta$}.

References: (1) \citet{esa97}; (2) \citet{ryab_etal98}; (3)
\citet{har97} and \citet*{jda99}; (4) \citet{har97}, \citet{jda99} and
\citet{wal_etal94}; (5) \citet{zav2007}; (6) \citet{ryab-etal96}.

\end{minipage}
\end{minipage}
\end{table*}

\subsection{Atomic data}

The oscillator strengths used in this analysis are given as \loggf\ in
Table~\ref{atomic}. For the $\lambda$4844.33 and
$\lambda$5292.22 lines, we obtained the oscillator strengths by using
the transition probabilities given by \citet*{ziel-etal02}. These authors
carried out a critical evaluation of experimental lifetimes in the
literature and measured their own branching ratios which they used to
calculate transition probabilities. They claimed uncertainties of $< 5$
per cent for some lines. \citet{ziel-etal02} did not include the
$\lambda$4603.03 line in their analysis and we obtained the oscillator
strength for this line by using the transition probability given by
\citet{gig-etal94} who used a pulsed arc as a plasma source and used a
method which enabled them to determine the plasma temperature and
transition probabilities simultaneously from relative intensities of
spectral lines. The lower-level excitation potentials, also listed in
Table~\ref{atomic}, are from \citet{rs88}. The LS designations are from 
\citet*{diroc-etal2000}.

We assumed the classical radiative damping constant
\mbox{$\Gamma_{\mathrm{R}} =$ 0.2223 x 10$^{16}/\lambda^{2}$ s$^{-1}$}
(where $\lambda$ is in \AA). This is a good approximation for these lines
as the typical lifetime of the upper level is about 10~ns. For the Van der
Waals damping we have followed the formulation by \citet{war67}.  At the
effective temperatures found in the HgMn stars, Van der Waals damping is
expected to make only a small contribution to the line broadening because
hydrogen is mostly ionized except at the coolest end of the temperature
range. For Stark broadening, we used the results of \citet{pd96} who
calculated theoretical values for the Stark widths of \Xeii\ lines and
compared them with measured values in the literature. Examination of their
paper shows that their theoretical widths are reasonably representative of
the measured widths for a wide range of experimental temperatures. The
variation with temperature was so small that we assumed the average
theoretical value for the Stark width (which we obtained from the listed
ratios of measured to theoretical values in \citealt{pd96}) at a
temperature of \mbox{$\teff\sim12000$ K} for all the HgMn programme stars.
We present the Stark widths in the form of
$\gamma_{\mathrm{s}}/n_{\mathrm{e}}$ as given in Table~\ref{atomic}. We
found that the most significant source of damping in our sample of HgMn
stars is $\gamma_{\mathrm{s}}$ which dominates by an order of magnitude
over radiative and Van der Waals damping.

\begin{table*}
\begin{minipage}{10cm}
\caption{Atomic data for \Xeii\ lines}
\label{atomic}
\begin{tabular}{lccrc}

\hline

LS designation & Wavelength & lower e.p. & \loggf\,$^a$ & 
$10^{6}$ $\gamma_{\mathrm{s}}/n_{\mathrm{e}}$\\
 & (\AA) & (eV) & & (cm$^{3}$s$^{-1}$)\\

\hline

$6s^{2}\mathrm{P}_{3/2}$ -- $6p^{2}\mathrm{P}_{3/2}$ & 4603.03 & 11.79 &
$-0.02$ & 6.58\\
$6s^{4}\mathrm{P}_{5/2}$ -- $6p^{4}\mathrm{D}_{7/2}$ & 4844.33 & 11.54 &
0.51 & 7.23\\
$6s^{4}\mathrm{P}_{5/2}$ -- $6p^{4}\mathrm{P}_{5/2}$ & 5292.22 & 11.54 &
0.38 & 6.46\\

\hline

\end{tabular}
\begin{minipage}{10cm}
$^a$Values of \loggf\ were calculated from the transition probabilities 
given by \citet{ziel-etal02} for $\lambda$4844.33 and $\lambda$5292.22, 
and \citet{gig-etal94} for $\lambda$4603.03.

\end{minipage}
\end{minipage}
\end{table*}

\subsection{Abundances and equivalent widths}
 
\begin{figure*}
\includegraphics[width=10cm,angle=270]{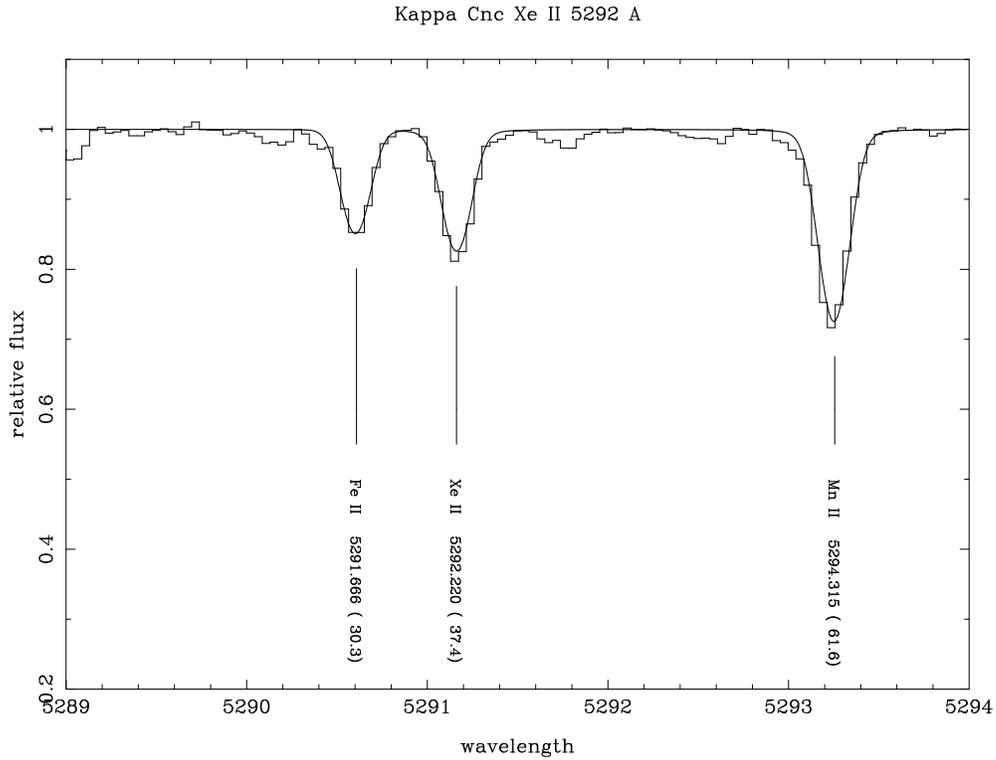}
\caption{\Xeii\ $\lambda$5292.22 in $\kappa$\,Cnc. The histogram
represents the observed spectrum. Solid line is synthesis from UCLSYN.
The wavelength shift is due to the synthesis including Doppler shifts in
both components of the binary. Figures in brackets following line
identifications are the observed equivalent widths in m\AA. Note 
that we did not attempt to identify all the weak lines in this 
spectrum.}
\label{kappaCnc-example}
\end{figure*}

\begin{figure*}
\includegraphics[width=10cm,angle=270]{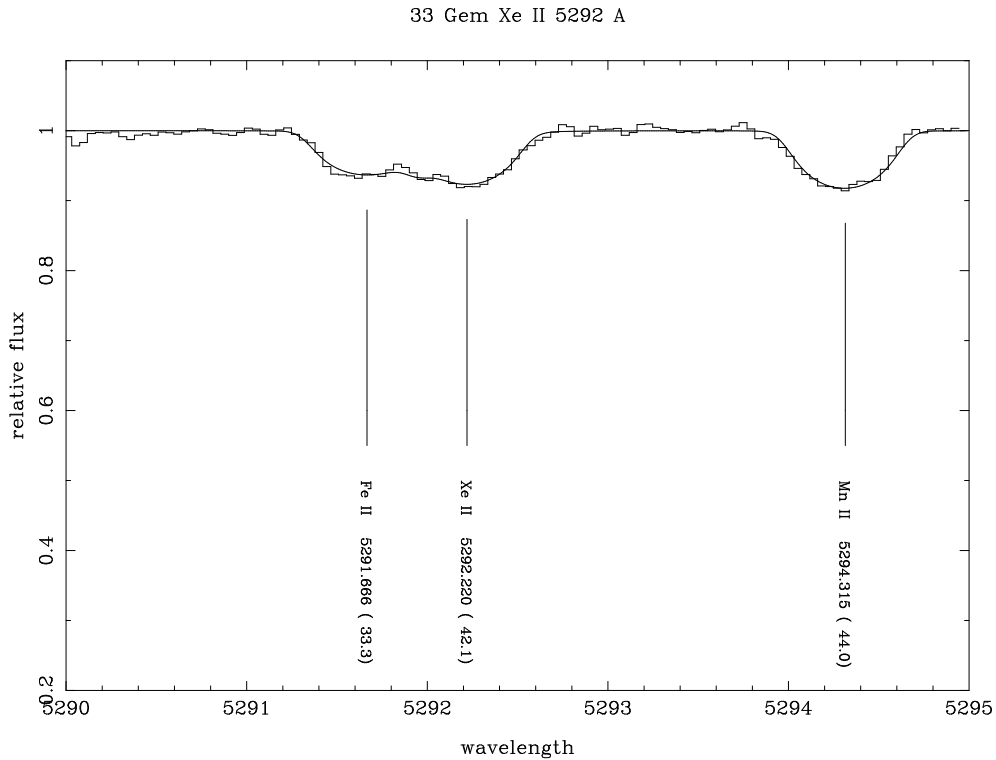}
\caption{\Xeii\ $\lambda$5292.22 in 33\,Gem. The higher rotational
velocity (\mbox{\vsini\ = 19.5 km\,s$^{-1}$}) creates some problems with
blends and therefore synthesis of the spectrum is used to fit the
blends.} 
\label{33Gem-example} 
\end{figure*}

Determination of abundances and equivalent widths was carried out using
the spectrum synthesis program UCLSYN in the LTE approximation. When the
\Xeii\ lines were sharp and unblended, equivalent widths were measured
interactively in UCLSYN and used in the program's exact curve of growth
analysis to obtain the abundance for that line. In cases where
rotational blending necessitated modeling of the spectra, we determined
abundances from minimized $\chi^{2}$ synthetic fits of the blended
line profile holding all parameters constant except the abundance of Xe.
In those cases, the equivalent width we give is that modeled from these
least-squares fits for the Xe lines. In most of the binaries the lines
were sufficiently unblended to allow an equivalent width to be measured
and an exact curve of growth analysis to be used. Some examples are
presented in Figure~\ref{kappaCnc-example} and
Figure~\ref{33Gem-example}.

\begin{figure*}
\includegraphics[width=10cm,angle=270]{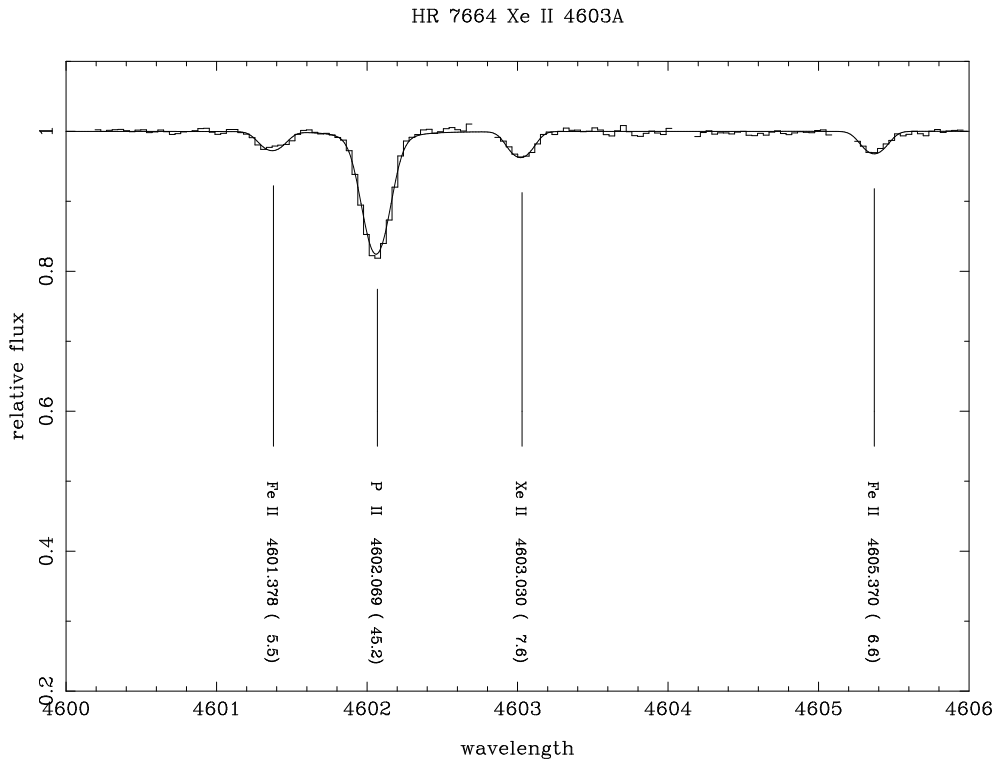}
\caption{\Xeii\ $\lambda$4603.03 in HR\,7664. Gaps in the observed 
spectrum indicate where cosmic rays or bad columns have been removed.}
\label{hr7664-weakxe}
\end{figure*}

Of the three \Xeii\ lines used, $\lambda$4603.03 was the weakest, but it
was generally detectable without any difficulty.
Figure~\ref{hr7664-weakxe} illustrates this line in the spectrum of HR
7664, where the nearest line is of very overabundant \Pii. 

\section{results}

The \Xeii\ equivalent widths and abundances for the HgMn stars are given
in Table~\ref{abunds}. Xenon was detected in 21 out of 22 HgMn stars in
the sample ($\phi$\,Her being the only exception). This is a somewhat
surprising result given the paucity of previous detections in the 
literature. The cosmic abundance of xenon from \citet{asp-etal05} is
2.27 on the scale where \mbox{log {\em A}\,(H) = 12.00.} The lower limit for
detection in our sample is of the order of \mbox{log {\em A}\,(Xe) =
5.2} depending on effective temperature and \vsini. In all the HgMn
stars where xenon was detected in this work, the mean \Xeii\ abundance
(Table~\ref{abunds}) is much greater than the cosmic abundance of xenon. 

\begin{table*}
\begin{minipage}{10cm}

\caption{\Xeii\ equivalent widths (m\AA) and abundances for HgMn 
programme stars, where $\log A$(H) = 12.00}

\label{abunds}

\begin{tabular}{lccccccl}

\hline

\Xeii\ Line & \multicolumn{2}{c}{$\lambda$4603.03} &
\multicolumn{2}{c}{$\lambda$4844.33} &
\multicolumn{2}{c}{$\lambda$5292.22} & Mean\\
Star & $W_\lambda$ & $\log A$ & $W_\lambda$ & $\log A$ & $W_\lambda$ &
$\log A$ & $\log A$\\

\hline

87\,Psc & $\leq2.6$ & $\leq5.44$ & 4.5: & 5.32: & 4.8: & 5.38: & 
5.35:$\pm 0.03$\\
53\,Tau & $\leq2.0$ & $\leq5.59$ & 3.1: & 5.60: & 3.3: & 5.57: & 
5.59:$\pm 0.02$\\
$\mu$\,Lep & 10.1 & 6.26 & 25.3 & 6.78 & 26.2 & 6.85 & 6.63$\pm 0.19$\\
HR\,1800$^a$ & 4.3 & 6.25 & 5.4 & 6.14 & 3.3 & 5.82 & 6.07$\pm 0.13$\\
33\,Gem & 34.2:$^\dagger$ & 7.24:$^\dagger$ & 38.5 & 6.98 &
42.1$^\dagger$ & 7.16$^\dagger$ & 7.10$\pm 0.07^{\ast\ast}$\\
HR\,2676 & $\leq1.7$ & $\leq5.04$ & 11.2 & 5.53 & $\leq7.5:^\dagger$ &
$\leq5.39:^\dagger$ & 5.53:\\
HR\,2844 & 7.7: & 5.94: & 29.9 & 6.76 & 20.4:$^\dagger$ &
6.36:$^\dagger$ & 6.46$\pm 0.20^{\ast\ast}$\\
$\kappa$\,Cnc$^a$ & 20.8 & 6.88 & 36.1 & 7.19 & 37.4 & 7.32 & 7.13$\pm 
0.13$\\
HR\,4072$^a$ & 3.9 & 6.17 & 6.6 & 6.29 & 4.9 & 6.06 & 6.17$\pm 0.07$\\
$\chi$\,Lup$^a$ & 3.6 & 6.25 & 5.7 & 6.38 & 4.5 & 6.15 & 6.26$\pm 
0.07$\\
$\iota$\,CrB$^a$ & $\ast$ & $\ast$ & 4.5 & 6.18 & 5.3 & 6.24 & 
6.21$\pm 0.03$\\
$\upsilon$\,Her & 2.9:$\ast$ & 5.60:$\ast$ & 8.9 & 5.94 & 7.5 & 5.85 & 
5.84$\pm 0.07^{\ast\ast}$\\
$\phi$\,Her$^a$ & $\leq1.2$ & $\leq5.39$ & $\leq1.2$ & $\leq5.16$ &
$\leq1.2$ & $\leq5.11$ & $\leq5.22\pm 0.09$\\
28\,Her & 2.8:$^\dagger$ & 5.81:$^\dagger$ & 6.8: & 6.11: & 5.9 & 6.00 & 
5.98$\pm 0.05^{\ast\ast}$\\
HR\,6997 & 24.6 & 6.62 & 41.8 & 6.78 & 45.2 & 6.93 & 6.78$\pm 0.09$\\
112\,Her$^a$ & 10.0 & 6.40 & 22.1 & 6.84 & 20.9 & 6.78 & 6.67$\pm 
0.14$\\
HR\,7143 & 4.0 & 5.82 & 10.8 & 6.15 & 9.1 & 6.01 & 5.99$\pm 0.10$\\
HR\,7361 & 21.6 & 6.71 & 35.4 & 6.93 & 41.1 & 7.24 & 6.96$\pm 0.15$\\
46\,Aql & 6.6:$\ast$ & 5.86:$\ast$ & 15.5 & 6.08 & 17.7: & 6.29: & 
6.08$\pm 0.08^{\ast\ast}$\\
HR\,7664 & 7.6 & 5.88 & 18.8 & 6.11 & 17.2 & 6.12 & 6.04$\pm 0.08$\\
HR\,7775 & 4.2:$^\dagger$ & 6.17:$^\dagger$ & 5.7 & 6.19 & 6.5 & 6.20 & 
6.19$\pm 0.01^{\ast\ast}$\\
$\beta$\,Scl & 9.0 & 6.26 & 16.1 & 6.42 & 19.2$^\dagger$ & 
6.61$^\dagger$ & 6.43$\pm 0.10$\\

\hline
\end{tabular}
\begin{minipage}{10cm}
$^a$Double-lined spectroscopic binary (visual binary in the case of
HR\,1800) with observed $W_\lambda$; abundance takes secondary dilution
into account. 

Notation ``$\ast$'' means bad column or cosmic ray hit.

Notation ``:'' means approximate.

Notation ``$\dagger$'' means that the line was blended.

Notation ``$\ast\ast$'' indicates that approximate values are 
given half-weight in the calculation of the mean $\log A$. In such 
cases, the error quoted is the error on the weighted mean. 

Note that, except for $\phi$\,Her, upper limits are not included when
calculating the mean $\log A$.

\end{minipage}
\end{minipage}
\end{table*}

\begin{figure}
\includegraphics[width=6.4cm,angle=270]{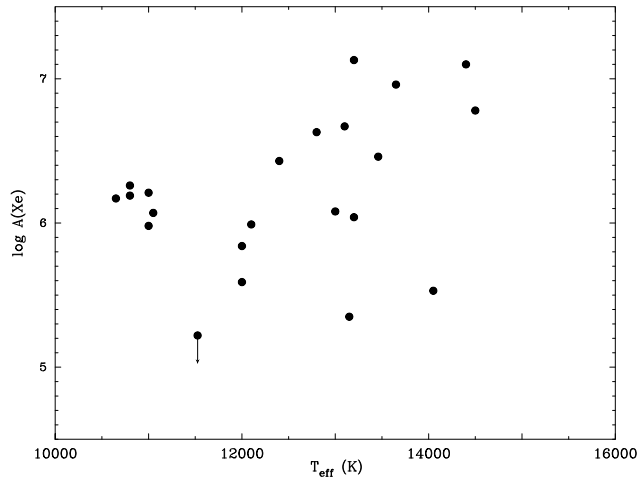}
\caption{Mean xenon abundances as a function of \teff\ in the HgMn
programme stars. Arrow indicates upper limit. Solar \mbox{log {\em 
A}\,(Xe) = 2.27.}}
\label{fig-abunds}
\end{figure}

In Figure~\ref{fig-abunds} the mean xenon abundances are plotted as a
function of \teff. There is a weak correlation but it is not
significant ($r = 0.35 \pm 0.24$). We found no significant correlation 
with surface gravity or \vsini.

We examined available spectra of normal late B stars in the \citet{sd93} 
list and searched for any lines at positions of the \Xeii\ lines. As 
expected, we found no sign of any absorption features corresponding to 
\Xeii\ in the normal late B stars.

At this stage, no calculations of the NLTE effects for \Xeii\ have been
attempted, and it is quite possible that this may modify the results
presented here. However, a vast overabundance of xenon in nearly all
HgMn stars seems an inescapable conclusion from this work.

\section{Error analysis}

The errors quoted on the mean abundances in Table~\ref{abunds} are formal
standard errors on the mean, weighted where appropriate. As the
calculation of such errors involves small-number statistics, these formal
errors are not especially reliable. We also considered the errors due to
uncertainties in the atmospheric parameters or observational data. To
address this, we have examined the effect on the abundances of
uncertainties in \teff, \logg, microturbulent velocity, Stark broadening
and equivalent width, by varying these parameters one at a time whilst
holding all other parameters constant. This was done for HR\,7143, one of
the weaker-lined stars in our sample, and HR\,7361 which is one of the
strong-lined stars in our sample.  The adopted uncertainties in the
parameters were: $\pm250$ K for \teff; $\pm0.10$ dex for \logg; $\pm0.5$
km\,s$^{-1}$ for microturbulent velocity; and $\pm50$ \% for
$\gamma_{\mathrm{s}}/n_{\mathrm{e}}$. In a star that rotates more rapidly,
the noise will affect more pixels than it would over a narrower line;
therefore we have adopted an error in equivalent-width of $\pm2$ m\AA\ for
HR\,7361 and $\pm1$ m\AA\ for HR\,7143. The results of this error analysis
are shown in Table~\ref{errors} where we list the absolute values of the
errors in $\log A$. Some of the errors were slightly asymmetric
and in such cases the mean absolute error in abundance is given. The
errors were added in quadrature to derive the total error given in the
last column of Table~\ref{errors}.

\begin{table*}
\begin{minipage}{17cm}

\caption{Typical absolute abundance errors resulting from uncertainties in
parameters or observational data}

\label{errors}
\begin{tabular}{llcccccc}

\hline

\Xeii\ Line & Star & \multicolumn{5}{c}{Error in log abundance for:} & 
Total\\
\cline{3-7}
 & & $\Delta$\teff $=\pm250$ K & $\Delta$\logg $=\pm0.10$
dex & $\Delta$\vturb $=\pm0.5$ km\,s$^{-1}$ & $\Delta
\gamma_{\mathrm{s}}/n_{\mathrm{e}} = \pm50$ \% & $\Delta W_\lambda$ & \\

\hline

$\lambda$4603.03 & HR\,7143 & 0.04 & 0.04 & 0.02 & 0.00 & 0.14 & 0.15\\
 & HR\,7361 & 0.04 & 0.03 & 0.04 & 0.02 & 0.10 & 0.12\\
$\lambda$4844.33 & HR\,7143 & 0.08 & 0.07 & 0.05 & 0.01 & 0.07 & 0.14\\
 & HR\,7361 & 0.04 & 0.04 & 0.07 & 0.07 & 0.08 & 0.14\\
$\lambda$5292.22 & HR\,7143 & 0.06 & 0.04 & 0.04 & 0.01 & 0.08 & 0.12\\
 & HR\,7361 & 0.04 & 0.02 & 0.07 & 0.09 & 0.08 & 0.15\\

\hline

\end{tabular}

\begin{minipage}{15.3cm}
For HR\,7143 ($\teff = 12100$ K), we adopted the estimate $\Delta
W_\lambda = \pm1$ m\AA; for HR\,7361 ($\teff = 13650$ K), \mbox{$\Delta
W_\lambda = \pm2$ m\AA}.

\end{minipage}
\end{minipage}
\end{table*}

\section{Discussion}

\subsection{The xenon enhancement in HgMn stars}

Xenon (\Xeii) is yet another of the spectra originally omitted from the
Revised Multiplet Tables by \citet{moore45} as `not of astrophysical
interest' yet subsequently found in the HgMn stars and other CP types (e.g.,
\Gaii,\ \Krii,\ \Ptii,\ \Auii,\ \Hgii,\ \Tlii\ and \Biii\ have also been
identified in stars).

The xenon enhancement in HgMn stars rivals that of other rare elements 
such as Hg \citep{s97,wl99} with many of the stars having Xe enhancement 
factors of \mbox{40--50\,000}. Such a large enhancement of abundance 
implies a strong upwards radiative acceleration on Xe over a wide range 
of \teff\ and \logg,\ and it will be interesting to see if this 
prediction is confirmed by future calculations, and to see whether Xe is 
expected to be stratified in a thin layer. Radiative diffusion processes 
must be dredging up xenon from deep inside the stellar envelope to get 
an enhancement of this size. As can be seen in Figure~\ref{fig-abunds}, 
the enhancements found in the cooler HgMn stars clustered around 
\mbox{11\,000 K} are all very similar at about 3.9 dex. At higher 
\teff,\ the enhancements show much more scatter, which is not an 
artifact of the measurement uncertainties but must be a real effect. The 
one star with an upper limit only, $\phi$\,Her, could itself have a 
considerable enhancement of Xe but yet be below the limit of 
detectability in the present work.

We looked at what we feel are the 10 best-observed stars in 
Table~\ref{abunds} to see if there were any systematic differences in 
results from the lines used. The mean difference between \mbox{log 
{\em{A}}} derived from $\lambda$4844 and $\lambda$5292 is \mbox{0.03 
$\pm$ 0.06 dex}, so these two lines give results that agree very well. 
The mean difference between abundances from $\lambda$4603 and the 
average of the other two lines is \mbox{-0.22 $\pm$ 0.08 dex}, which 
seems obviously significant. The most likely explanation is an error in 
one or more of the adopted oscillator strengths. However, this does not 
affect any of the essential conclusions in this work.

The rarity of \Xeii\ identifications in previous work is explicable by 
the relative weakness of its lines in the photographic blue region, 
where many observations were done only below 4600\AA. It is easy to 
understand why the lines might be overlooked in some earlier work of 
this sort -- and in at least one case, a strong line of another element 
masks the presence of a \Xeii\ line, as discussed below.

\subsection{A comment on strontium abundances}

In the course of an investigation of strontium abundances in HgMn stars
\citep*{ddp08} it was noticed that \Srii\ $\lambda$4215.519 apparently
persisted even when \Srii\ $\lambda$4077.714 was completely undetectable.
An example of this is seen in the \Srii\ results of \citet{ryab-etal96},
where $\lambda$4215.519 has a measured equivalent width of 6.5m\AA\ in
112\,Her. We also found a similar strength for this line, but a wavelength
slightly shifted.  We were unable to detect \Srii\ $\lambda$4077.714 at
all ($\leq$0.6m\AA), nor did \citet{ryab-etal96} report seeing this line.

The solution to this anomaly is our identification of a blend of \Srii\ with
\Xeii\ $\lambda$4215.60 \citep{hp87}, which has an estimated strength of
4m\AA\ in 112\,Her from the abundance in Table~\ref{abunds} and an
astrophysical \loggf\ of $-1.06$, based on the Lick observations of the
ultra-sharp-lined, Sr-deficient, ``mild'' HgMn star 46\,Aql
\citep{cow80,sd93}. The remaining strength of this feature in 112\,Her is
probably due to weak lines of \hbox{Mn\,{\sc ii}} and \hbox{Cr\,{\sc ii}}.
Strontium is at least 1.0 dex below solar abundance in both 46\,Aql and
112\,Her.

\section{Acknowledgements}

The authors are grateful for the assistance of 2006 MSc
project student Mr.~Alex Dyer and his detective work on the lines of
strontium. Observations obtained at Lick Observatory were supported by
generous allotments of Guest Observer time by the Director, J. Miller,
and financial support for travel was provided by the UK Particle Physics
and Astronomy Research Council through its PATT grant to UCL. We
gratefully acknowledge the use of data from the UVES Paranal Observatory
Project (ESO DDT Program ID 266.D-5655). We are grateful to the anonymous 
referee for helpful comments that led to improvements in this paper.


\begin{thebibliography}{}

\bibitem[\protect\citeauthoryear{Adelman}{1987}]{adel87}
Adelman S.J., 1987, MNRAS, 228, 573

\bibitem[\protect\citeauthoryear{Adelman}{1988a}]{adel88a}
Adelman S.J., 1988a, MNRAS, 235, 749

\bibitem[\protect\citeauthoryear{Adelman}{1988b}]{adel88b}
Adelman S.J., 1988b, MNRAS, 235, 763

\bibitem[\protect\citeauthoryear{Adelman}{1989}]{adel89}
Adelman S.J., 1989, MNRAS, 239, 487

\bibitem[\protect\citeauthoryear{Adelman}{1992}]{adel92}
Adelman S.J., 1992, MNRAS, 258, 167

\bibitem[\protect\citeauthoryear{Adelman \& Fuhr}{1985}]{adel_fuhr85}
Adelman S.J., Fuhr J. R., 1985, A\&A, 152, 434

\bibitem[\protect\citeauthoryear{Adelman, Philip \& Adelman}{Adelman et
al.}{1996}]{adel-etal96} Adelman S.J., Philip A.G.D., Adelman C.J.,
1996, MNRAS, 282, 953

\bibitem[\protect\citeauthoryear{Allen}{1998}]{al98} Allen C. S., 1998, 
PhD thesis, Univ. London 
(http://www.ulo.ucl.ac.uk/ulo\_comms/80/index.html)

\bibitem[\protect\citeauthoryear{Andersen, Jaschek \& Cowley}{Andersen
et al.}{1984}]{andersen_etal84} Andersen J., Jaschek M., Cowley C. R.,
1984, A\&A, 132, 354

\bibitem[\protect\citeauthoryear{Asplund, Grevesse \& Sauval}{Asplund et
al.}{2005}]{asp-etal05} Asplund M., Grevesse N., Sauval A.J., 2005, in
Barnes, III, T.G., Bash F.N., eds, Asp Conf. Ser. Vol. 336, Cosmic
abundance as records of stellar evolution and nucleosynthesis. Astron.
Soc. Pac., San Francisco, p.25

\bibitem[\protect\citeauthoryear{Bagnulo et al.}{2003}]{bag-etal03} 
Bagnulo S., Jehin E., Ledoux C., Cabanac R., Melo C., Gilmozzi R., 
2003, Messenger, 114, 10

\bibitem[\protect\citeauthoryear{Bidelman}{1962}]{bid62}
Bidelman W., 1962, Sky \& Telescope, 23, 140 

\bibitem[\protect\citeauthoryear{Bohlender, Dworetsky \& Jomaron}
{Bohlender et al.}{1998}]{boh_etal98} Bohlender D. A., Dworetsky M. M.,
Jomaron C. M., 1998, ApJ, 504, 533

\bibitem[\protect\citeauthoryear{Churchill}{1995}]{church95} Churchill
C.  W., 1995, Lick Obs. Tech. Rep. No. 74

\bibitem[\protect\citeauthoryear{Cowley}{1980}]{cow80} Cowley C. R., 1980, 
PASP, 92, 159

\bibitem[\protect\citeauthoryear{Di Rocco, Iriarte \& Pomarico}{Di Rocco
et al.}{2000}]{diroc-etal2000} Di Rocco H. O., Iriarte D. I., Pomarico
J. A., 2000, Eur. Phys. J. D, 10, 19

\bibitem[\protect\citeauthoryear{Dworetsky \& Budaj}{2000}]{db2000} 
Dworetsky M. M., Budaj J., 2000, MNRAS, 318, 1264

\bibitem[\protect\citeauthoryear{Dworetsky, Dyer \& Persaud}{Dworetsky et
al.}{2008}]{ddp08} Dworetsky M. M., Dyer A., Persaud J. L., 2008, Contrib.
Astron. Obs. Skalnat\'{e} Pleso, in press

\bibitem[\protect\citeauthoryear{Dworetsky, Jomaron \&
Smith}{Dworetsky et al.}{1998}]{djs98} Dworetsky M. M., Jomaron C. M.,
Smith C. A., 1998, A\&A, 333, 665

\bibitem[\protect\citeauthoryear{ESA}{1997}]{esa97} ESA, 1997, The 
Hipparcos and Tycho Catalogues. ESA-SP 1200. ESA Publications 
Division, Noordwijk

\bibitem[\protect\citeauthoryear{Gigosos et al.}{1994}]{gig-etal94}
Gigosos M. A., Mar S., P\'{e}rez C., de la Rosa I., 1994, Phys. Rev. E,
49, 1575

\bibitem[\protect\citeauthoryear{Guthrie}{1985}]{guth85} Guthrie B. N. G., 
1985, MNRAS, 216, 1

\bibitem[\protect\citeauthoryear{Hansen \& Persson}{1987}]{hp87} Hansen 
J. E., Persson W., 1987, Phys. Scripta, 36, 602

\bibitem[\protect\citeauthoryear{Harman}{1997}]{har97} Harman D. J., 
1997, MSci Project Report, Univ. College London

\bibitem[\protect\citeauthoryear{Jomaron, Dworetsky \& Allen}{Jomaron 
et al.}{1999}]{jda99} Jomaron C. M., Dworetsky M. M., Allen C. S., 
1999, MNRAS, 303, 555

\bibitem[\protect\citeauthoryear{Kurucz}{1993}]{kurucz93} Kurucz R. L., 
1993, ATLAS9 Stellar Atmosphere Programs and \mbox{2-km\,s$^{-1}$} Grid 
(CD-ROM 13)

\bibitem[\protect\citeauthoryear{Misch}{1997}]{mis97} Misch A., 1997, 
User's Guide to the Hamilton Echelle Spectrometer 
(http://www.ucolick.org/$\sim$tony/instruments/hamspec/
hamspec\_index.html)

\bibitem[\protect\citeauthoryear{Moore}{1945}]{moore45} Moore C. E., 
1945, A Multiplet Table of Astrophysical Interest (Revised Edition), 
Contrib. Princeton Univ. Obs. No. 20

\bibitem[\protect\citeauthoryear{Popovi\'{c} \& 
Dimitrijevi\'{c}}{1996}]{pd96} Popovi\'{c} L. \v{C}., Dimitrijevi\'{c} 
M. S., 1996, A\&AS, 116, 359

\bibitem[\protect\citeauthoryear{Ryabchikova \& Smirnov}{1988}]{rs88}
Ryabchikova T.A., Smirnov J.M., 1988, Astron. Tsirk., 1534, 21

\bibitem[\protect\citeauthoryear{Ryabchikova, Zakharova \& Adelman}
{Ryabchikova et al.}{1996}]{ryab-etal96}
Ryabchikova T.A., Zakharova L.A., Adelman S.J., 1996, MNRAS, 283, 1115

\bibitem[\protect\citeauthoryear{Ryabchikova et 
al.}{1998}]{ryab_etal98} Ryabchikova T., Kotchoukhov O., Galazutdinov 
F., Musaev F., Adelman S. J., 1998, Contrib. Astron. Obs. Skalnat\'{e} 
Pleso, 27, 258

\bibitem[\protect\citeauthoryear{Sadakane et al.}{2001}]{sada-etal01}
Sadakane K., et al., 2001, PASJ, 53, 1223

\bibitem[\protect\citeauthoryear{Smith}{1992}]{s92} Smith K.C., 1992, 
PhD thesis, Univ. London

\bibitem[\protect\citeauthoryear{Smith}{1993}]{s93} Smith K.C., 1993,
A\&A, 276, 393

\bibitem[\protect\citeauthoryear{Smith}{1994}]{s94} Smith K.C., 1994,
A\&A, 291, 521

\bibitem[\protect\citeauthoryear{Smith}{1996}]{s96} Smith K.C., 1996,
A\&A, 305, 902

\bibitem[\protect\citeauthoryear{Smith}{1997}]{s97} Smith K.C., 1997,
A\&A, 319, 928

\bibitem[\protect\citeauthoryear{Smith \& Dworetsky}{1988}]{sd88} Smith 
K.C., Dworetsky M.M., 1988, in Adelman S.J., Lanz T., eds, Elemental 
Abundance Analyses, 99. Institut d'Astronomie de l'Univ. de Lausanne, 
Switzerland, p.32

\bibitem[\protect\citeauthoryear{Smith \& Dworetsky}{1993}]{sd93}
Smith K.C., Dworetsky M.M., 1993, A\&A, 274, 335

\bibitem[\protect\citeauthoryear{Valdes}{1990}]{val90} Valdes F., 1990, 
The IRAF APEXTRACT Package (ftp://iraf.tuc.noao.edu/iraf/docs/apex.ps)

\bibitem[\protect\citeauthoryear{Vogt}{1987}]{vogt87} Vogt S., 1987, 
PASP, 99, 1214

\bibitem[\protect\citeauthoryear{Wahlgren}{2004}]{wal04} Wahlgren G. M., 
2004, in Zverko J., \v{Z}i\v{z}\v{n}ovsk\'{y} J., Adelman S. J., Weiss 
W. W., eds, Proc. IAU Symp. 224, The A-star Puzzle, CUP, p. 291

\bibitem[\protect\citeauthoryear{Wahlgren, Adelman \& Robinson}
{Wahlgren et al.}{1994}]{wal_etal94} Wahlgren G. M., Adelman S. J., 
Robinson R. D., 1994, ApJ, 434, 349

\bibitem[\protect\citeauthoryear{Warner}{1967}]{war67} Warner B., 1967, 
MNRAS, 136, 381.

\bibitem[\protect\citeauthoryear{Woolf \& Lambert}{1999}]{wl99} Woolf
V. M., Lambert D. L., 1999, ApJ, 521, 414

\bibitem[\protect\citeauthoryear{Zavala et al.}{2007}]{zav2007} Zavala 
R.T. et al., 2007, ApJ, 655, 1046

\bibitem[\protect\citeauthoryear{Zieli\'{n}ska, Bratasz \&
Dzier\.{z}\c{e}ga}{Zieli\'{n}ska et al.}{2002}]{ziel-etal02} 
Zieli\'{n}ska S., Bratasz \L.,
Dzier\.{z}\c{e}ga K., 2002, Phys. Scripta, 66, 454


\end{thebibliography}
\end{document}